\documentclass[aps,prl,twocolumn,groupedaddress]{revtex4}
\usepackage{graphicx}

\usepackage[figuresright]{rotating}

\begin{document}

\title{First Study for the Pentaquark Potential in SU(3) Lattice QCD}

\author{Fumiko~Okiharu}
\affiliation{Department of Physics, Nihon University, 1-8 Kanda-Surugadai, Chiyoda, Tokyo 101, Japan}

\author{Hideo~Suganuma}
\affiliation{Faculty of Science, Tokyo Institute of Technology, 2-12-1 Ohokayama, Tokyo 152-8551, Japan}

\author{Toru~T.~Takahashi}
\affiliation{Yukawa Institute for Theoretical Physics, Kyoto University, Kitashirakawa, Sakyo, Kyoto 606-8502, Japan}

\date{\today}

\begin{abstract}
The static penta-quark (5Q) potential $V_{\rm 5Q}$ is studied 
in SU(3) lattice QCD with $16^3\times 32$ and $\beta$=6.0 at the quenched level. 
From the 5Q Wilson loop, $V_{\rm 5Q}$ is calculated in a gauge-invariant manner, 
with the smearing method to enhance the ground-state component.
$V_{\rm 5Q}$ is well described by the OGE plus multi-Y Ansatz: 
a sum of the OGE Coulomb term and the multi-Y-type linear term 
proportional to the minimal total length of the flux-tube linking the five quarks. 
Comparing with ${\rm Q \bar Q}$ and 3Q potentials, 
we find a universality of the string tension, 
$\sigma_{\rm Q \bar Q} \simeq \sigma_{\rm 3Q} \simeq \sigma_{\rm 5Q}$, and the OGE result for Coulomb coefficients.
\end{abstract}


\maketitle


The inter-quark force is one of the elementary quantities for the study of 
the multi-quark system in the quark model.
As for baryons, our group recently studied the three-quark (3Q) potential $V_{\rm 3Q}$ in detail with lattice QCD, 
and clarified that it obeys the Coulomb plus Y-type linear potential\cite{TS010203}.
However, no one knows the inter-quark force from QCD in the exotic multi-quark system such as tetra-quark mesons (QQ-$\bar {\rm Q}\bar{\rm Q}$),  
penta-quark baryons (4Q-$\bar{\rm Q}$), dibaryons (6Q) and so on.

Very recently, an exotic anti-strange baryon $\Theta^+(1540)$ with $S=+1$ was 
experimentally discovered at SPring-8 (LEPS), 
and was confirmed by ITEP(DIANA), JLab(CLAS) and ELSA(SAPHIR)\cite{Theta1540}.
The $\Theta^+(1540)$ was theoretically predicted in the Skyrme model\cite{DPP97}, and is 
regarded as a penta-quark (5Q) baryon of $\rm u^2d^2\bar s$ in the valence-quark picture.
Another 5Q baryon $\Xi^{--}(1862)$ was found by CERN(NA49)\cite{NA49}, 
and also an anti-charmed 5Q baryon $\Theta_c(3099)$ was found by HERA(H1)\cite{H1}.

Accordingly, many theoretical studies\cite{O04} have been done for the 5Q baryon using  various approaches such as 
lattice QCD\cite{CFKK03,STOI04}, the constituent quark model\cite{KL03}, 
the diquark model\cite{JW03}, the QCD sum rule\cite{Z03SDO04}, 
the flux-tube model\cite{SZ04}, the string theory\cite{BKST04} and so on. 
However, there are several puzzling problems on the $\Theta^+(1540)$:
its mass seems to be rather small and its decay width is extremely narrow\cite{Theta1540}.
To solve them, one encounters the many-body problem of quarks, and therefore 
it is quite desired to clarify the inter-quark force in the multi-quark system based on QCD.

In this paper, motivated by the recent discovery of the penta-quark baryons, 
we perform the first study of the static penta-quark (5Q) potential $V_{\rm 5Q}$,  
i.e., the inter-quark force in the 5Q system, 
in SU(3) lattice QCD with $\beta$=6.0 and $16^3 \times$ 32 at the quenched level.
Note that the lattice QCD result of $V_{\rm 5Q}$ presents a key information 
in modeling the multi-quark system based on QCD. 

For the penta-quark system, 
we investigate the QQ-$\bar {\rm Q}$-QQ type configuration with the two ``QQ clusters" belonging to the {\bf 3*} 
representation of the SU(3) color as shown in Fig.1, since this type of the 5Q configuration is expected to have a small energy and 
seems to be natural as a realistic candidate of the $\Theta^+(1540)$.
Indeed, in the perturbative sense, an attractive (repulsive) force acts between two quarks, when  
their total SU(3) color belongs to {\bf 3*} ({\bf 6}). 
Therefore, the nearest QQ cluster tends to form {\bf 3*} rather than {\bf 6} in  the low-lying 5Q system, 
which leads to the {\bf 3*}-diquark model\cite{JW03}.
\begin{figure}[h]
\begin{center}
\includegraphics[scale=0.3]{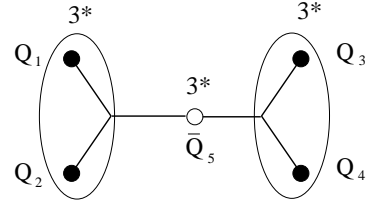}
\caption{\label{fig:1}
The QQ-$\bar {\rm Q}$-QQ type configuration for the penta-quark system.
The two QQ clusters belong to the {\bf 3}* representation of the color SU(3).
}
\end{center}
\end{figure}


\begin{figure}[h]
\begin{center}
\includegraphics[scale=0.4]{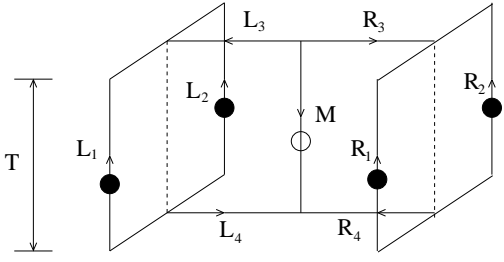}
\caption{\label{fig:2}
The penta-quark (5Q) Wilson loop $W_{\rm 5Q}$ for the 5Q potential $V_{\rm 5Q}$.
The contours $M, L_i, R_i (i=3,4)$ are line-like and $L_i, R_i (i=1,2)$ are staple-like.
The 5Q gauge-invariant state is generated at $t=0$ and is annihilated at $t=T$ with 
the five quarks (4Q-$\bar{\rm Q}$) being spatially fixed in ${\bf R}^3$ for $0 < t < T$. 
}
\end{center}
\end{figure}

Similar to the derivation of the Q-${\bar {\rm Q}}$ (3Q) potential from the (3Q) Wilson loop, 
the 5Q static potential $V_{\rm 5Q}$ can be calculated with the 5Q Wilson loop $W_{\rm 5Q}$, 
which is defined in a gauge-invariant manner as shown in Fig.2.

We define the 5Q Wilson loop $W_{\rm 5Q}$ \cite{STOI04} as  
\begin{eqnarray}
W_{\rm 5Q}\equiv \frac1{3!} \epsilon^{abc} \epsilon^{a'b'c'}M^{aa'}
(\tilde L_3\tilde L_{12}\tilde L_4)^{bb'}(\tilde R_3\tilde R_{12}\tilde R_4)^{cc'}, 
\end{eqnarray}
where $\tilde M, \tilde L_i, \tilde R_i (i=1,2,3,4)$ are given by 
\begin{eqnarray}
\tilde M, \tilde L_i, \tilde R_i \equiv P\exp\{ig \int_{M, L_i, R_i} dx^{\mu}A_{\mu}(x)\} \in {\rm SU(3)}_c.
\end{eqnarray}
As shown in Fig.2, $\tilde M, \tilde L_i, \tilde R_i (i=3,4)$ are line-like variables and 
$\tilde L_i, \tilde R_i (i=1,2)$ are staple-like variables.
Here, $\tilde L_{12}, \tilde R_{12} \in {\rm SU(3)}_c$ are defined as   
\begin{eqnarray}
\tilde L_{12}^{a'a}\equiv \frac12 \epsilon^{abc} \epsilon^{a'b'c'}\tilde L_1^{bb'} \tilde L_2^{cc'},
~
\tilde R_{12}^{a'a}\equiv \frac12 \epsilon^{abc} \epsilon^{a'b'c'}\tilde R_1^{bb'} \tilde R_2^{cc'}. 
\label{R12}
\end{eqnarray}
Note that the 5Q Wilson loop $W_{\rm 5Q}$ is gauge invariant, and its gauge invariance is owing to  
the nontrivial assignment of the color indices of $\tilde L_{12}$ and $\tilde R_{12}$ in 
Eq.(\ref{R12}).
(Recall that the ``two quark lines" combining into the 3* representation 
correspond to an ``antiquark line" as the color current.)


In principle, the ground-state 5Q potential $V_{\rm 5Q}$ is obtained from 
the 5Q Wilson loop $\langle W_{\rm 5Q}\rangle$ as 
$V_{\rm 5Q}=-\lim_{T \rightarrow \infty} \frac1T \ln \langle W_{\rm 5Q}\rangle$.
However, the practical lattice calculation is done with a finite region of $T$, 
where excited-state contributions remain.
For the accurate measurement of $V_{\rm 5Q}$ in lattice QCD, 
we use the gauge-covariant smearing method\cite{TS010203} to enhance 
the ground-state component of the 5Q state in the 5Q Wilson loop.

The smearing is known to be a powerful method for the accurate 
measurement of the Q-$\bar {\rm Q}$ and the 3Q potentials\cite{TS010203}, 
and is expressed as the iterative replacement of 
the spatial link variables $U_i(s)$ ($i$=1,2,3) 
by the obscured link variables $\bar U_i(s)\in {\rm SU}(3)_c$ 
which maximizes
${\rm Re} \,{\rm tr} \,\,\{\bar U_i^{\dagger}(s) V_i(s)\}$ with
\begin{eqnarray}
V_i(s)\equiv 
\alpha U_i(s)+\sum_{j \ne i} \sum_{\pm} 
\{ U_{\pm j}(s)U_i(s\pm \hat j)U_{\pm j}^\dagger (s+\hat i) \} 
\end{eqnarray}
with the simplified notation of $U_{-j}\equiv U^\dagger_{j}(s-\hat j)$. 
We here adopt $\alpha=2.3$ and the iteration number $N_{\rm smr}=40$, which lead to 
a large enhancement of the ground-state component in the 5Q Wilson loop at $\beta$=6.0. 


Now, we proceed the actual lattice QCD calculation for the 5Q potential $V_{\rm 5Q}$ \cite{STOI04}. 
We generate 150 gauge configurations using SU(3)$_c$ lattice QCD 
with the standard action with $\beta=6.0$ and $16^{3} \times 32$ at the quenched level.
The gauge configurations are taken every 500 sweeps 
after a thermalization of 5000 sweeps using the pseudo-heat-bath algorithm.
The lattice spacing $a$ is estimated as $a \simeq 0.104{\rm fm}$ 
from the string tension $\sigma$=0.89 GeV/fm in the Q-$\bar{\rm Q}$ potential $V_{\rm Q \bar{Q}}$ \cite{STOI04}.

As for the 5Q configuration, 
we consider the QQ-$\bar {\rm Q}$-QQ type configuration as shown in Fig.1.
Note that the multi-quark system including four or more quarks can take a three-dimensional shape, 
while the Q$\bar {\rm Q}$ and the 3Q systems can take only planar configuration\cite{STOI04,SZ04}. 
Then, we investigate both the planar 5Q configuration as shown in Fig.3 and 
the twisted 5Q configuration as shown in Fig.4.
In this paper, we take $d_1=d_2=d_3=d_4\equiv d$, and present the lattice QCD result of $V_{\rm 5Q}$ 
in terms of $(d, h_1,h_2)$. 

\begin{figure}[h]
\begin{center}
\includegraphics[scale=0.3]{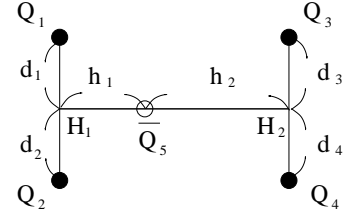}
\caption{\label{fig:3}
A planar configuration of 
the penta-quark system.
$ {\rm Q}_1{\rm Q}_2$ is parallel to ${\rm Q}_3{\rm Q}_4$, and 
${\rm H}_1{\rm H}_2$ is perpendicular to 
${\rm Q}_1{\rm Q}_2$ and ${\rm Q}_3{\rm Q}_4$.
Here, we take $d_1=d_2=d_3=d_4\equiv d$.
}
\end{center}
\end{figure}

\begin{figure}[h]
\begin{center}
\vspace{-1cm}
\includegraphics[scale=0.3]{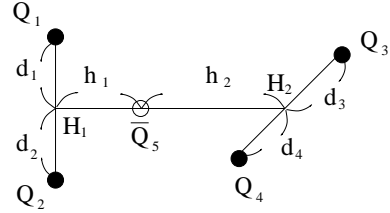}
\caption{\label{fig:4}
A twisted configuration of 
the penta-quark system.
$ {\rm Q}_1{\rm Q}_2$ is perpendicular to ${\rm Q}_3{\rm Q}_4$, and 
${\rm H}_1{\rm H}_2$ is perpendicular to 
${\rm Q}_1{\rm Q}_2$ and ${\rm Q}_3{\rm Q}_4$.
Here, we take $d_1=d_2=d_3=d_4\equiv d$.
}
\end{center}
\end{figure}

For these types of 5Q configurations, 
we calculate the 5Q potential $V_{\rm 5Q}$ from 
the 5Q Wilson loop $\langle W_{\rm 5Q}\rangle$ using the smearing method.
Owing to the smearing, the ground-state component is largely enhanced,  
and therefore the 5Q Wilson loop $\langle W_{\rm 5Q} \rangle$ 
composed with the smeared link variable exhibits 
a single-exponential behavior as 
$\langle W_{\rm 5Q} \rangle \simeq e^{-V_{\rm 5Q}T}$ 
even for a small value of $T$. 
Then, for each 5Q configuration, we extract $V_{\rm 5Q}$
from the least squares fit with the single-exponential form
\begin{equation}
\langle W_{\rm 5Q}\rangle =\bar{C}e^{-V_{\rm 5Q}T}
\label{expfit}
\end{equation}
in the range of $T_{\rm min}\leq T\leq T_{\rm max}$ listed in Table~I.
The prefactor $\bar C$ physically means the ground-state overlap, and 
$\bar C\simeq 1$ corresponds to the quasi-ground-state.
Here, we choose the fit range of $T$ such that the stability of the
``effective mass'' $V(T)\equiv \ln\{\langle W_{\rm 5Q}(T) \rangle /
\langle W_{\rm 5Q}(T+1)\rangle\}$
is observed in the range of $T_{\rm min}\leq T\leq T_{\rm max}-1$.
For the lattice calculation of $\langle W_{\rm 5Q}\rangle$,
we use the translational and the rotational symmetries on lattices.

\begin{table}[h]
\begin{center}
\newcommand{\m}{\hphantom{$-$}}
\newcommand{\cc}[1]{\multicolumn{1}{c}{#1}}
\renewcommand{\tabcolsep}{0.3pc} 
\renewcommand{\arraystretch}{1} 
\caption{\label{tab1}
Lattice QCD results for the penta-quark potential $V_{\rm 5Q}$ 
for the planar 5Q configuration labeled by $(d,h_1,h_2)$ as shown in Fig.3. 
We list also the ground-state overlap $\bar C$, the fit range of $T$ and 
the theoretical form $V_{\rm 5Q}^{\rm theor}$ of the OGE plus multi-Y Ansatz (\ref{theorform}) 
with ($A_{\rm 5Q}$,$\sigma_{\rm 5Q}$) fixed to be ($A_{\rm 3Q}$,$\sigma_{\rm 3Q}$) in  
$V_{\rm 3Q}$ in Ref.\cite{TS010203}.
All the data are measured in the lattice unit.
}
\small
\begin{tabular}{ccccc}
\hline\hline
$(d,h_1,h_2)$ 
& $\rm V_{\rm 5Q}$ & $\bar C$ & $T_{\rm min}$-$T_{\rm max}$ & $V_{\rm 5Q}^{\rm theor}$
\\ 
\hline
(1,1,1) & 1.4452(11)  & 0.9539(21)  & 2-7 & 1.4433 \\
(1,1,2) & 1.5409(13)  & 0.9506(25)  & 2-8 & 1.5414 \\ 
(1,1,3) & 1.6177(19)  & 0.9512(33)  & 2-7 & 1.6146 \\  
(1,1,4) & 1.6793(20)  & 0.9431(35)  & 2-7 & 1.6767 \\  
(1,1,5) & 1.7381(23)  & 0.9394(40)  & 2-6 & 1.7332 \\  
(1,1,6) & 1.7918(28)  & 0.9311(49)  & 2-6 & 1.7866 \\  
(1,1,7) & 1.8441(31)  & 0.9232(56)  & 2-6 & 1.8380 \\  
(1,2,2) & 1.6314(17)  & 0.9503(29)  & 2-6 & 1.6322 \\  
(1,2,3) & 1.7011(20)  & 0.9427(34)  & 2-5 & 1.7021 \\  
(1,2,4) & 1.7680(25)  & 0.9478(43)  & 2-6 & 1.7623 \\  
(1,2,5) & 1.8190(29)  & 0.9297(50)  & 2-6 & 1.8177 \\  
(1,2,6) & 1.8717(33)  & 0.9205(57)  & 2-5 & 1.8704 \\  
(1,3,3) & 1.7723(24)  & 0.9405(39)  & 2-4 & 1.7702 \\  
(1,3,4) & 1.8336(28)  & 0.9351(49)  & 2-7 & 1.8293 \\  
(1,3,5) & 1.8913(33)  & 0.9320(62)  & 2-5 & 1.8839 \\  
(1,4,4) & 1.8939(31)  & 0.9293(53)  & 2-4 & 1.8877 \\
(2,1,1) & 1.7531(23)  & 0.9393(40)  & 2-5 & 1.7515 \\ 
(2,2,2) & 1.8803(31)  & 0.9292(54)  & 2-6 & 1.8887 \\    
(2,3,3) & 2.0030(37)  & 0.9284(64)  & 2-5 & 2.0098 \\  
(2,4,4) & 2.1122(49)  & 0.9116(91)  & 2-5 & 2.1211 \\ 
(3,1,1) & 1.9734(37)  & 0.9138(66)  & 2-5 & 1.9850 \\ 
(3,2,2) & 2.0811(45)  & 0.9070(75)  & 2-5 & 2.0942 \\
(3,3,3) & 2.1886(53)  & 0.9003(92)  & 2-4 & 2.2047 \\ 
(3,4,4) & 2.3043(68)  & 0.9084(113) & 2-5 & 2.3105 \\ 
(4,1,1) & 2.1697(60)  & 0.8948(100) & 2-5 & 2.1958 \\
(4,2,2) & 2.2734(60)  & 0.8890(100) & 2-5 & 2.2829 \\
(4,3,3) & 2.3657(73)  & 0.8606(120) & 2-4 & 2.3864 \\  
(4,4,4) & 2.4706(104) & 0.8534(164)& 2-5 & 2.4884 \\ 
\hline\hline
\end{tabular}
\end{center}
\end{table}

\begin{table}[h]
\begin{center}
\newcommand{\m}{\hphantom{$-$}}
\newcommand{\cc}[1]{\multicolumn{1}{c}{#1}}
\renewcommand{\tabcolsep}{0.3pc} 
\renewcommand{\arraystretch}{1} 
\caption{\label{tab2}
Lattice QCD results for the penta-quark potential $V_{\rm 5Q}$
for the twisted 5Q configuration labeled by $(d,h_1,h_2)$ as shown in Fig.4. 
The notations are the same in Table~I.
}
\small
\begin{tabular}{ccccc}
\hline\hline
$(d,h_1,h_2)$ 
& $V_{\rm 5Q}$ & $\bar C$ & $T_{\rm min}$-$T_{\rm max}$ & $V_{\rm 5Q}^{\rm theor}$
\\ 
\hline
(1,1,1) & 1.4476(23)  & 0.9378(64)  & 3-8 & 1.4458 \\
(1,1,2) & 1.5438(14)  & 0.9528(25)  & 2-8 & 1.5419 \\ 
(1,1,3) & 1.6155(17)  & 0.9459(31)  & 2-6 & 1.6148 \\  
(1,1,4) & 1.6767(21)  & 0.9370(37)  & 2-6 & 1.6767 \\  
(1,1,5) & 1.7365(22)  & 0.9357(42)  & 2-5 & 1.7332 \\  
(1,1,6) & 1.7912(26)  & 0.9297(42)  & 2-4 & 1.7866 \\  
(1,1,7) & 1.8337(88)  & 0.8933(231) & 3-5 & 1.8380 \\  
(1,2,2) & 1.6302(16)  & 0.9472(29)  & 2-8 & 1.6324 \\  
(1,2,3) & 1.7022(18)  & 0.9445(32)  & 2-4 & 1.7022 \\  
(1,2,4) & 1.7657(25)  & 0.9427(44)  & 2-5 & 1.7624 \\  
(1,2,5) & 1.8232(30)  & 0.9385(51)  & 2-7 & 1.8177 \\  
(1,2,6) & 1.8728(32)  & 0.9230(58)  & 2-8 & 1.8704 \\  
(1,3,3) & 1.7710(24)  & 0.9376(42)  & 2-6 & 1.7702 \\  
(1,3,4) & 1.8326(27)  & 0.9335(46)  & 2-4 & 1.8293 \\  
(1,3,5) & 1.8952(32)  & 0.9394(58)  & 2-7 & 1.8839 \\  
(1,4,4) & 1.8950(30)  & 0.9315(52)  & 2-4 & 1.8877 \\
(2,1,1) & 1.7735(23)  & 0.9377(41)  & 2-6 & 1.7615 \\ 
(2,2,2) & 1.8832(30)  & 0.9279(54)  & 2-7 & 1.8899 \\    
(2,3,3) & 2.0011(37)  & 0.9233(64)  & 2-5 & 2.0100 \\  
(2,4,4) & 2.1155(49)  & 0.9167(85)  & 2-6 & 2.1212 \\ 
(3,1,1) & 2.0049(37)  & 0.9032(67)  & 2-5 & 2.0008 \\ 
(3,2,2) & 2.0873(38)  & 0.8987(69)  & 2-4 & 2.0973 \\
(3,3,3) & 2.1870(54)  & 0.8912(92)  & 2-6 & 2.2055 \\ 
(3,4,4) & 2.3021(68)  & 0.9019(113) & 2-5 & 2.3108 \\ 
(4,1,1) & 2.2141(65)  & 0.8741(107) & 2-6 & 2.2155 \\
(4,2,2) & 2.2874(64)  & 0.8768(107) & 2-5 & 2.2879 \\
(4,3,3) & 2.3715(70)  & 0.8577(118) & 2-4 & 2.3880 \\  
(4,4,4) & 2.4680(94)  & 0.8459(149) & 2-4 & 2.4890 \\ 
\hline\hline
\end{tabular}\\
\end{center}
\end{table}

For 56 different patterns of the 5Q configurations as shown in Figs.3 and 4, 
we present the lattice QCD data for the 5Q potential $V_{\rm 5Q}$ 
together with the ground-state overlap $\bar{C}$ 
in Table~I and II. 
The statistical errors are estimated 
with the jackknife method. 
We find a large ground-state overlap as $\bar{C} > 0.85$ 
for almost all 5Q configurations.


Next, we consider the theoretical form of the 5Q potential $V_{\rm 5Q}$.
The lattice QCD studies\cite{TS010203} at the quenched level show that 
the Q-$\bar {\rm Q}$ potential $V_{\rm Q \bar{Q}}$ takes a form of 
\begin{equation}
V_{\rm Q \bar{Q}}(r)=-\frac{A_{\rm Q \bar{Q}}}{r}
+\sigma_{\rm Q \bar{Q}} r+C_{\rm Q \bar{Q}},  
\end{equation}
and the 3Q potential $V_{\rm 3Q}$ takes a form of 
\begin{equation}
V_{\rm 3Q}=-A_{\rm 3Q}\sum_{i<j}\frac1{|{\bf r}_i-{\bf r}_j|}
+\sigma_{\rm 3Q} L_{\rm min}+C_{\rm 3Q}, 
\end{equation}
where $L_{\rm min}$ denotes the minimal value of 
total length of color flux tubes linking the three quarks.
In fact, both $V_{\rm Q \bar{Q}}$ and $V_{\rm 3Q}$
are described by a sum of the short-distance one-gluon-exchange (OGE) 
result and the long-distance flux-tube result\cite{TS010203,IBSS03}. 

For the static penta-quark (5Q) system, we find that 
the lattice QCD results are well described by the OGE plus multi-Y Ansatz:  
a sum of the OGE Coulomb term and the multi-Y type linear term \cite{STOI04},
\begin{eqnarray}
&&V_{\rm 5Q}
=\frac{g^2}{4\pi} \sum_{i<j} \frac{T^a_i T^a_j}{|{\bf r}_i-{\bf r}_j|}+\sigma_{\rm 5Q} L_{\rm min}+C_{\rm 5Q} \nonumber \\
&=&-A_{\rm 5Q}\{ ( \frac1{r_{12}}  + \frac1{r_{34}}) 
+\frac12(\frac1{r_{15}} +\frac1{r_{25}} +\frac1{r_{35}} +\frac1{r_{45}}) \nonumber \\
&+&\frac14(\frac1{r_{13}} +\frac1{r_{14}} +\frac1{r_{23}} +\frac1{r_{24}}) \}
+\sigma_{\rm 5Q} L_{\rm min}+C_{\rm 5Q}
\label{theorform}
\end{eqnarray}
with 
$r_{ij}\equiv|{\bf r}_i-{\bf r}_j|$ and $i$th quark location ${\bf r}_i$ in Fig.1.
Here, $L_{\rm min}$ is the minimal length of the flux-tube linking five quarks as shown in Fig.1.
(For the extreme case, e.g., $d>\sqrt{3}h_1$, we here assume that the flux-tube is formed as the two straight lines 
on ${\rm Q}_1{\rm Q}_5$ and ${\rm Q}_2{\rm Q}_5$, considering the color combination,  
although there may appear several possibilities as the ``flip-flop".)

Note that there appear three kinds of Coulomb coefficients ($A_{\rm 5Q}$, $\frac12 A_{\rm 5Q}$, $\frac14 A_{\rm 5Q}$) in the penta-quark system, 
while only one Coulomb coefficient, $A_{\rm Q\bar {\rm Q}}$ or $A_{\rm 3Q}$, appears in the Q$\bar{\rm Q}$ or the 3Q system. 
Here, the Coulomb coefficient $A_{\rm 5Q}$ in Eq.(\ref{theorform}) corresponds to $A_{\rm 3Q}$ or $\frac12 A_{\rm Q\bar Q}$ in terms of the OGE result.

We add in Table~I and II the theoretical form $V_{\rm 5Q}^{\rm theor}$ 
of the OGE plus multi-Y Ansatz (\ref{theorform}) 
with ($A_{\rm 5Q}$,$\sigma_{\rm 5Q}$) fixed to be ($A_{\rm 3Q}$,$\sigma_{\rm 3Q}$) 
in the 3Q potential $V_{\rm 3Q}$ obtained in Ref.\cite{TS010203}, i.e., 
$A_{\rm 5Q}=A_{\rm 3Q}\simeq 0.1366$, 
$\sigma_{\rm 5Q}=\sigma_{\rm 3Q}\simeq 0.046a^{-2}$ and $C_{\rm 5Q}\simeq 1.58 a^{-1}$.
(Note that there is no adjustable parameter for the theoretical form $V_{\rm 5Q}^{\rm theor}$ 
besides an irrelevant constant $C_{\rm 5Q}$, 
since $A_{\rm 5Q}$ and $\sigma_{\rm 5Q}$ are fixed 
to be $A_{\rm 3Q}$ and $\sigma_{\rm 3Q}$, respectively.)
Thus, the 5Q potential $V_{\rm 5Q}$ is found to be well described by 
the OGE Coulomb plus multi-Y-type linear potential.

We show in Fig.5 typical examples of the lattice QCD data for the penta-quark potential $V_{\rm 5Q}$.
The symbols denote the lattice data, and the curves denote the theoretical form  
of the OGE plus multi-Y Ansatz 
with ($A_{\rm 5Q}$,$\sigma_{\rm 5Q}$) fixed to be ($A_{\rm 3Q}$,$\sigma_{\rm 3Q}$). 
One finds a good agreement between the lattice QCD data and the theoretical curves.

\begin{figure}[h]
\begin{center}
\rotatebox{-90}{\includegraphics[scale=0.3]{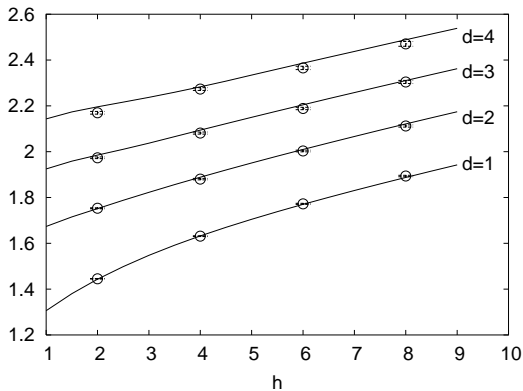}}
\caption{\label{fig:5}
Lattice QCD results of the penta-quark potential $V_{\rm 5Q}$ 
for the planar 5Q configuration with $h_1=h_2\equiv h$ in Fig.3 in the lattice unit. 
Each 5Q system is labeled by $d$ and $h$. 
The symbols denote the lattice data, and the curves 
the theoretical form of the OGE plus multi-Y Ansatz. 
}
\end{center}
\end{figure}

Note that the planar and the twisted 5Q configurations with the same $(d,h_1,h_2)$ are almost degenerate, 
although the energy of the planar one is slightly smaller.
In terms of the OGE plus multi-Y Ansatz, 
the only energy difference between the two states originates from a small difference of the Coulomb interaction 
between ${\rm Q}_i (i=1,2)$ and ${\rm Q}_j(j=3,4)$, where the Coulomb coefficient is reduced as 
$\frac14 A_{\rm 5Q} (\simeq \frac18 A_{\rm Q\bar Q})$. 
Then, no special configuration is favored in the 5Q system in terms of the energy. 
This fact also indicates that the 5Q system is unstable against the twisted motion of the two QQ clusters as shown in Fig.4. 
In fact, general 5Q systems tend to take a three-dimensional configuration \cite{STOI04,SZ04} in terms of the entropy.

From the comparison with the Q$\bar {\rm Q}$ 
and the 3Q potentials, the universality of the string tension 
and the OGE result are found among Q$\bar {\rm Q}$, 3Q and 5Q systems as 
\begin{eqnarray}
\sigma_{\rm Q\bar{\rm Q}}\simeq \sigma_{\rm 3Q} \simeq \sigma_{\rm 5Q}, \qquad
\frac12A_{\rm Q\bar{\rm Q}}\simeq A_{\rm 3Q} \simeq A_{\rm 5Q}.
\end{eqnarray} 
This result supports the flux-tube picture on the confinement mechanism 
even for the multi-quark system \cite{STOI04}.

To conclude, we have performed the first study of the penta-quark potential in lattice QCD, and have found that 
the 5Q potential is well reproduced by the OGE Coulomb plus multi-Y-type linear potential.

H.S. was supported in part by a Grant for Scientific Research 
(No.16540236) from the Ministry of Education, 
Culture, Science and Technology, Japan. 
T.T.T. was supported by the Japan Society for the Promotion of Science.
The lattice QCD Monte Carlo calculations have been performed on NEC-SX5 at Osaka University.

\end{document}